\documentclass[iop,apj,revtex4]{emulateapj}
\usepackage{graphicx}
\usepackage{amsmath,amssymb}

\begin{document}
\title{Proton heating by pick-up ion driven cyclotron waves in the outer heliosphere: Hybrid
expanding box simulations}

\author{Petr Hellinger\altaffilmark{1,2}}
\email{petr.hellinger@asu.cas.cz}
 \author{Pavel M. Tr\'avn\'\i\v cek\altaffilmark{1--3}}

\altaffiltext{1}{Astronomical Institute, CAS,
Bocni II/1401, CZ-14100 Prague, Czech Republic
}
\altaffiltext{2}{Institute of Atmospheric Physics, CAS,
Bocni II/1401, CZ-14100 Prague, Czech Republic
}
\altaffiltext{3}{Space Sciences Laboratory, UCB, Berkeley, USA.}

\begin{abstract}
Using one-dimensional hybrid expanding box model we investigate
properties of the solar wind in the outer heliosphere. We assume a proton-electron plasma 
with a strictly transverse ambient magnetic field and, beside the expansion, we take into account 
influence of a continuous injection of cold pick-up protons through the charge-exchange process
between the solar wind protons and hydrogen of interstellar origin.
The injected cold pick-up protons form a ring distribution function that 
rapidly becomes unstable and generate Alfv\'en cyclotron waves. 
The Alfv\'en cyclotron waves scatter pick-up protons to a spherical shell
distribution function that thickens over that time owing to the 
expansion-driven cooling. The Alfv\'en cyclotron waves heat
solar wind protons in the perpendicular direction (with respect to the ambient magnetic field)
 through the cyclotron resonance. At later
times, the Alfv\'en cyclotron waves become parametrically unstable
and the generated ion acoustic waves heat protons in the parallel direction through the
Landau resonance. The resulting heating of 
the solar wind protons is efficient on the expansion time scale.
\end{abstract}

\pacs{?}
\maketitle

\section{Introduction}

Properties of the solar wind plasma in the outer heliosphere are
importantly influenced by inelastic collisions between ions and neutrals
of interstellar origin \citep{burlal96,zankal09,rist09}. 
The solar wind ions and interstellar neutrals interact through
the charge-exchange process. The neutrals (mostly hydrogen) become ionized and
the interaction between the newly created, pick-up ions and the ambient solar
wind plasma leads to a deceleration of the solar wind 
and has likely an important effect on ion thermal energetics \citep{richal95,wangal00b,richal08}.
The nonthermal pick-up ions play likely an important role 
at the termination shock \citep{burlal08,deckal08}.
The actual form of the pick-up ion distribution function just upstream from
the termination shock importantly influences the shock properties and ion acceleration
\citep{gide10,masc14,yangal15}.

A pick-up ion dominated distant solar wind can be to some extend described 
using a macroscopic multi-fluid approach
\citep{zankal14}. Global multi-fluid simulations show that
the interaction between solar wind and interstellar neutrals has
an important effect on the global properties of the outer heliosphere
\cite[e.g.,][]{usmaal12}. 
However, the highly nonthermal newly born pick-up ion population leads naturally
to plasma instabilities where strongly kinetic processes such as
the cyclotron and Landau resonances are expected \citep{gary93}. 
Depending on the orientation of the background magnetic field,
the pick-up have different distribution functions \citep{zaca00}.
A ring distribution function is generated by injection perpendicular 
to the ambient solar wind. Such a distribution has an effective perpendicular
temperature anisotropy and may generate Alfv\'en cyclotron waves
(or mirror instabilities, etc.); the generated waves
scatter the ring pick-up protons to a spherical shell-like velocity distribution function
\citep{leip87,wiza94}
 and may heat directly
the solar wind protons through the cyclotron instability  \citep{grayal96,richal96}.
Injection parallel to the ambient magnetic field leads to several beam-type instabilities \citep{daga98}.
For a general orientation of the ambient magnetic field a ring-beam distribution function
is generated and both the temperature anisotropy and the differential velocity could be
a source of free energy for kinetic instabilities \citep{vahe15}.
The pick-up ion driven instabilities are active in the solar wind as indicated
by in situ observation of the generated waves 
\citep{joycal10,cannal14a,cannal14b,aggaal16}.

The linear and nonlinear studies of pick-up ion driven instabilities usually
assume a homogeneous plasma that is at odds with ubiquitous turbulent fluctuations
observed in the distant solar wind
\citep{fratal16}. In situ observations of enhanced wave activity 
\citep{joycal10,cannal14a,cannal14b,aggaal16} 
indicate that the pick-up ion driven instabilities are active in the solar wind.
This is supported by
 direct hybrid simulations of \cite{hellal15} showing that kinetic instabilities may coexist
with turbulence.
The wave activity generated by pick-up ions
may be a local source of turbulence 
as assumed in phenomenological transport 
models of turbulence
\citep{zankal96,mattal99,isenal03,smital06,isenal10,adhial15};
the enhanced level of turbulence leads to enhanced cascade
and heating rates. However,
connection between turbulence and kinetic instabilities
driven by non-thermal particle velocity distribution functions
is far from being understood.

In this paper we investigate long-time evolution of the expanding solar wind plasma
under the effect of a continuous injection of pick-up protons.
Section~\ref{heb} describes the simulation method, the hybrid expanding box model.
Section~\ref{results} presents the main simulation results. In section~\ref{discussion}
we discuss the simulation results.

\section{Expanding box model}

\label{heb}
To study the response of the plasma to a slow expansion we use
the expanding box model \citep{grapal93,liewal01}
implemented to the
hybrid code \citep{matt94,hellal03}.
In the hybrid expanding box (HEB) model
the expansion is described as an external force where a
constant solar wind radial velocity $v_{sw}$ is assumed. The radial distance $R$
is then
\begin{equation}
R = R_0 + v_{sw} t = R_0 \left(1+ \frac{t}{t_{e0}}\right)
\label{distancetime}
\end{equation}
where $R_0$ is an initial radial distance and $t_{e}=R_0/v_{sw}$
 is the characteristic (initial) expansion time.
The initial radial distance $R_0$ is one of the free parameters in the
model. Here we assume a strictly transverse magnetic field and
$R_0\sim 10$~au.
Transverse scales (with respect to the radial direction)
 of a small portion of plasma, co-moving with the solar
wind velocity, increase  with time
$\propto (1+t/t_{e0})$.
The expanding box uses these co-moving coordinates,
replacing the spatial dependence by the temporal one (see Equation~(\ref{distancetime})).
The physical transverse scales of the simulation
box increase with time \cite[see ][for a detailed description
of the code]{hetr05}
and the standard periodic boundary conditions are used.

The kinetic model uses the hybrid approximation, i.e., electrons are
considered as a massless, charge neutralizing fluid,
with a constant temperature; ions are
described by a particle-in-cell model and are advanced
 by a Boris' scheme that requires
the fields to be known at half time steps ahead of the
particle velocities. This is achieved by advancing the current density to
this time step with only one computational pass
through the particle data at each time step \citep{matt94}.
The characteristic spatial and temporal units used in the model
are $c/\omega_{p\mathrm{p}0}$ and $1/\omega_{c\mathrm{p}0}$ respectively,
where $c$ is the speed of light, $\omega_{p\mathrm{p}0} = ({n_{\mathrm{p}0}
e^2}/{m_\mathrm{p}\epsilon_0})^{1/2}$ is the initial proton plasma
frequency, and $\omega_{c\mathrm{p}0} = {eB_{0}}/{m_\mathrm{p}}$ is the initial
proton gyrofrequency
($B_{0}$: the initial magnitude of the ambient magnetic
 field $\boldsymbol{B}_0$,
$n_{\mathrm{p}0}$ is the initial proton density,
$e$ and $m_\mathrm{p}$ are the proton electric
charge and mass, respectively; finally,
$\epsilon_0$ is the dielectric permittivity of
vacuum).
We use the spatial resolution
$\Delta x=  0.25 c/\omega_{p\mathrm{p}0}$, and there are initially 131,072 particles per cell
for the solar wind protons.
Fields and moments are defined on a 1-D (periodic) grid with
 $8192$ points. Protons are advances using
 a time step $\Delta t=0.05/\omega_{c\mathrm{p}0}$,
while the magnetic field $\boldsymbol{B}$
is advanced with a smaller time step $\Delta t_B = \Delta t/10$.
The initial ambient magnetic field is directed along the (transverse) $x$ direction,
parallel to the ambient magnetic field
$\boldsymbol{B}_{0}=(B_{0},0,0)$, and we impose a
continuous expansion in $x$ and $z$ directions with
the characteristic time $t_e=10^4 /\omega_{c\mathrm{p}0}$. 
The radial direction is in the $y$ direction.
 The expansion leads
to a decrease of the density as $(1+{t}/{t_e})^{-2}$ whereas the magnitude 
of the magnetic field decreases as  $(1+{t}/{t_e})^{-1}$ for
the strictly transverse magnetic field.

In the simulation pick-up protons are continuously injected \cite[cf.][]{coweal08}.
The injection mechanism is the charge exchange between interstellar neutrals
and the solar wind protons. For the sake of simplicity we assume that the 
 charge exchange is energy/velocity independent.
For the loss rate of the incident solar wind protons with the distribution function $f_\mathrm{p}$ we have 
\begin{equation}
 \left(\frac{\mathrm{d} f_\mathrm{p}}{\mathrm{d} t}\right)_{cx} = - \nu_{cx} f_\mathrm{p}
\label{cx}
\end{equation}
where the charge-exchange frequency $\nu_{cx}$ is taken
to be $\nu_{cx}=10^{-6} \omega_{c\mathrm{p}0}$. The characteristic charge-exchange
time $t_{cx}=1/\nu_{cx}$ is 100 times longer than the expansion time $t_e$.
The solar wind protons that are lost through the charge-exchange process are removed from the simulation (but
their velocities are stored)
and replaced by cold pick-up protons with the mean radial velocity $v_R=-10 v_A \sim -v_{sw}$.
The  charge-exchange process, Equation~(\ref{cx}), leads to the decrease
of the solar wind proton number density (with respect to the electron one) 
as $n_\mathrm{p}/n_\mathrm{e}\propto \exp(-\nu_\mathrm{cx}t)$
(note that at the same time the particle number densities decrease owing to the expansion).
At the end of the simulation at $8t_e$ there is about 8 \% of pick-up protons.
The injection of pick-up protons also decelerates the solar wind plasma. At
the end of the simulation the solar wind protons are decelerated by about $0.75 v_A$.

\section{Simulation results}
\label{results}
The chosen initial conditions are stable with respect to kinetic instabilities.
The continuous injection of pick-up ions and the expansion tend to change the plasma properties
creating a free energy for kinetic instabilities.
Let us investigate the evolution of the HEB simulation starting with the wave activity.

\begin{figure}[htb]
\centerline{\includegraphics[width=8cm]{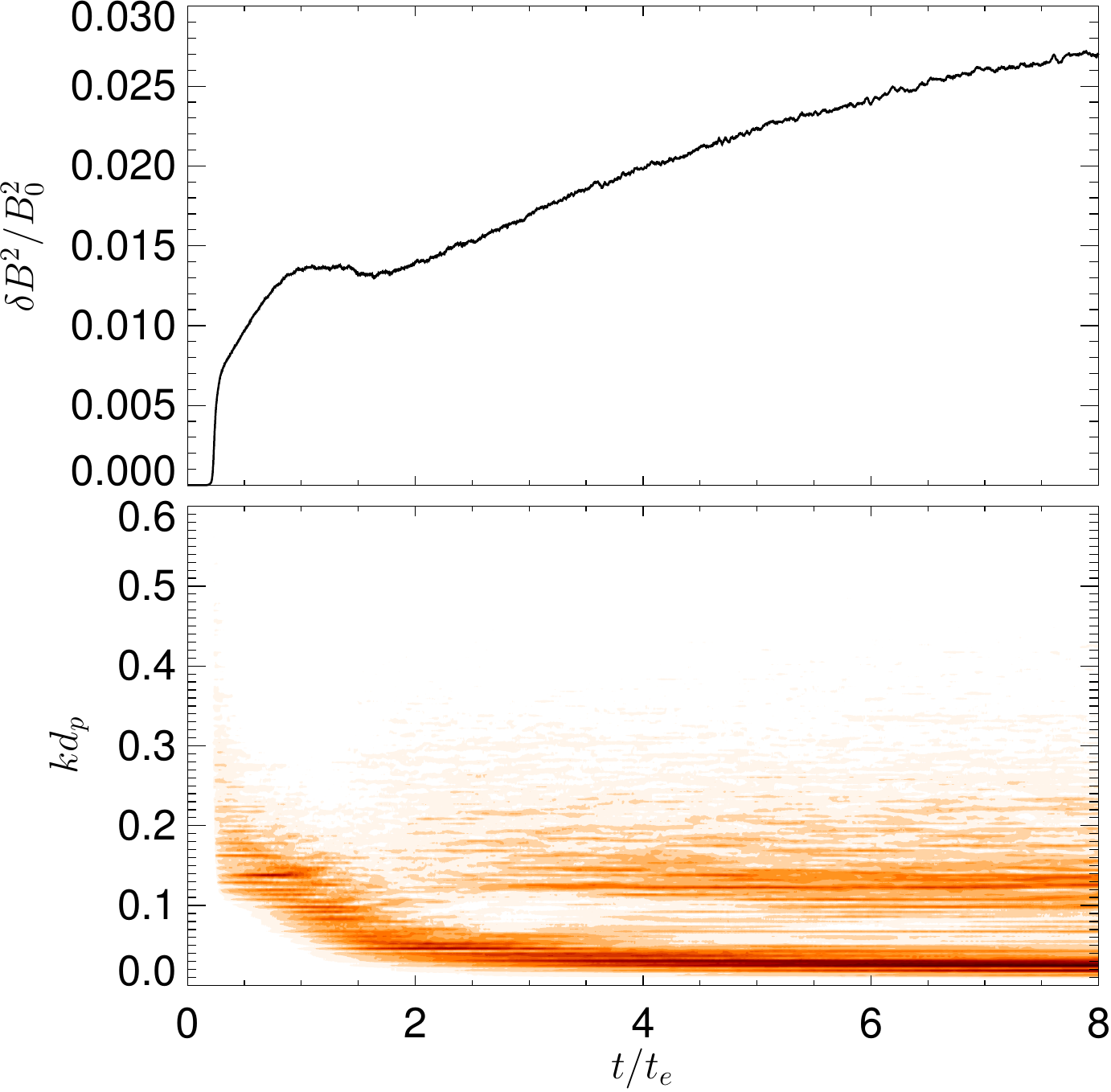}}
\caption{Evolution of fluctuating magnetic field:
(top) $\delta B^2$ as a function of $t$; (bottom) color scale plot
of $\delta B^2$ as a function of $t$ and $k$.
\label{db}
}
\end{figure}

Figure~\ref{db} shows an evolution of the fluctuating
magnetic field (normalized to the ambient magnetic field)
$\delta B^2/B_0^2$ as a function of time (top panel).
The bottom panel shows the color scale plot
of the fluctuating
magnetic field as a function of time and the wave vector $k$.
Figure~\ref{db} shows that $\delta B/B_0$ 
and grows exponentially  for 
$ 0.1 t_e \lesssim t \lesssim 0.2 t_e$.
After the saturation $\delta B$ 
grows secularly, i.e., about linearly in time, 
\cite[cf.,][]{mattal06,rosial11,kunzal14}.
Between $ t_e \lesssim t \lesssim 2 t_e$
there is another change of behavior, $\delta B/B_0$
remains about constant. For $t \gtrsim 2 t_e$
$\delta B/B_0$ grows again in a secular manner.
Bottom panel of Figure~\ref{db} shows that 
waves are initially generated with $0.1 \lesssim k d_\mathrm{p} \lesssim 0.2$.
The fluctuating wave energy then shifts to smaller wave vectors (longer
wavelengths) $k d_\mathrm{p} \sim 0.05$.  For $t \gtrsim 2 t_e$
there appears
a secondary population of wave modes with $0.1 \lesssim k d_\mathrm{p} \lesssim 0.16$.

\begin{figure}[htb]
\centerline{\includegraphics[width=8cm]{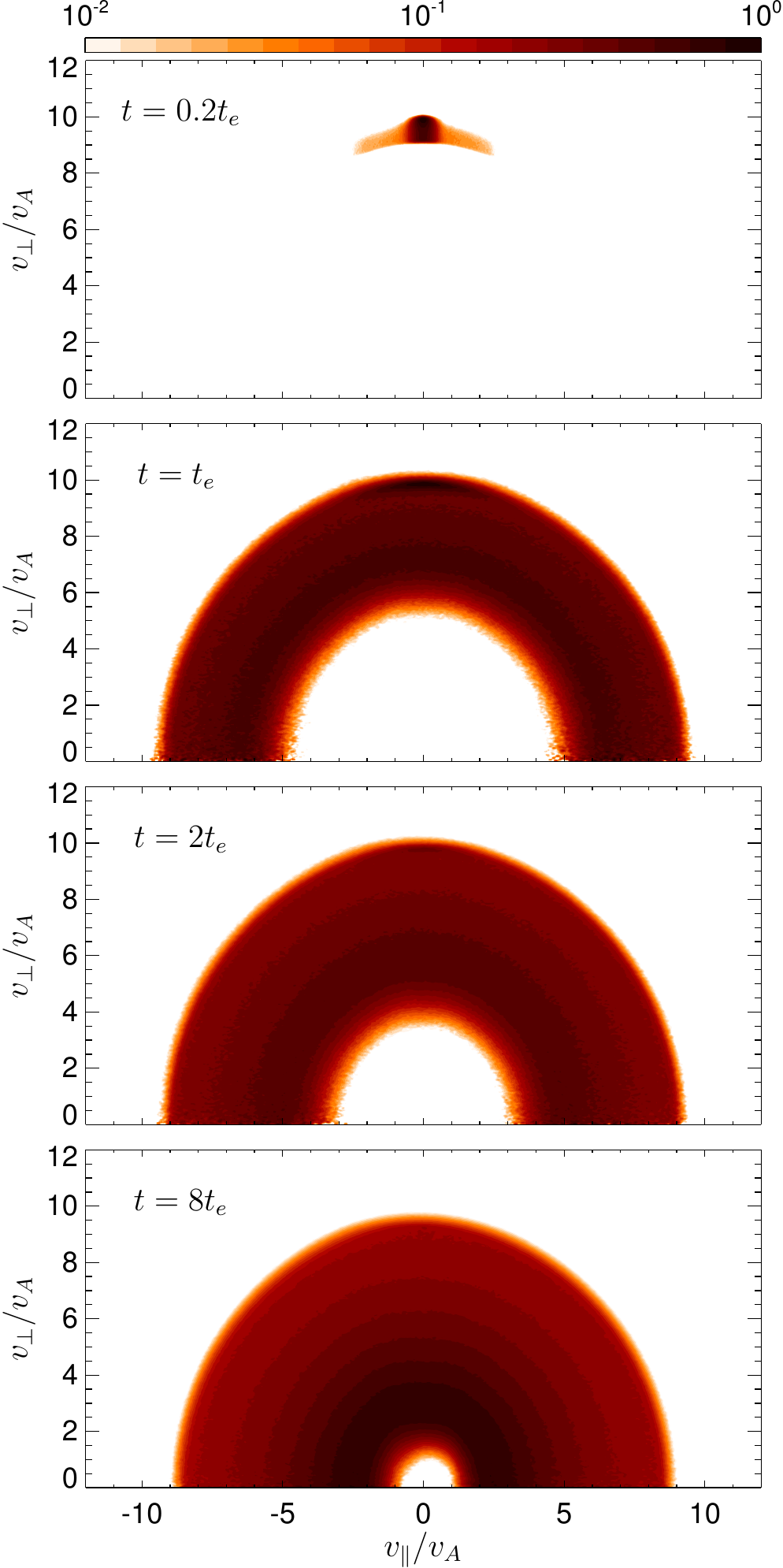}}
\caption{Evolution of the pick-up proton velocity distribution function (normalized to
its maximum value) as a
function of $v_\|$ and $v_\perp$
at $t=0.2t_e$, $t_e$, $2 t_e$, and $8 t_e$ (from top to bottom). The color scale is shown
on the top.
\label{evolvdfpui}
}
\end{figure}

The observed wave activity is generated by the continuously injected pick-up protons
that initially form a ring velocity distribution function \citep{vahe15}. The ring velocity distribution function
has a strong effective perpendicular temperature anisotropy that leads
to generation of Alfv\'en ion cyclotron (AIC) waves \citep{grayal96} which in turn diffuse
the ring ions.
Figure~\ref{evolvdfpui} shows the evolution of the pick-up proton velocity distribution functions.
At $t=0.2t_e$, i.e., around the saturation of the initial exponential growth
of the magnetic fluctuations 
the pick-up proton velocity distribution function takes shape of a
partially diffused ring that is expected for the saturated level of a weak ring instability \cite[cf.,][]{floral10}.
The pick-up proton velocity distribution function exhibit a strong concentration
of the newly born ions at the injection position, $v_\|\sim0$ and $v_\perp\sim 10 v_A$.
At $t=t_e$ the  pick-up proton velocity distribution exhibit a thick spherical shell velocity distribution function
as well as diffused enhancement around the injection region $v_\|\sim0$ and $v_\perp\sim 10 v_A$.
For a cold ring a thin  spherical shell is expected 
but the expansion leads to an anisotropic
cooling that transports the diffused pick-up protons to the regions with smaller velocities. 
At $t=2t_e$ the pick-up protons exhibit a shell velocity distribution function
further thickened owing to the expansion driven cooling.
This trend continues till $t=8t_e$, we observe no qualitative changes
between $t=2t_e$ and $t=8t_e$.

\begin{figure}[htb]
\centerline{\includegraphics[width=8cm]{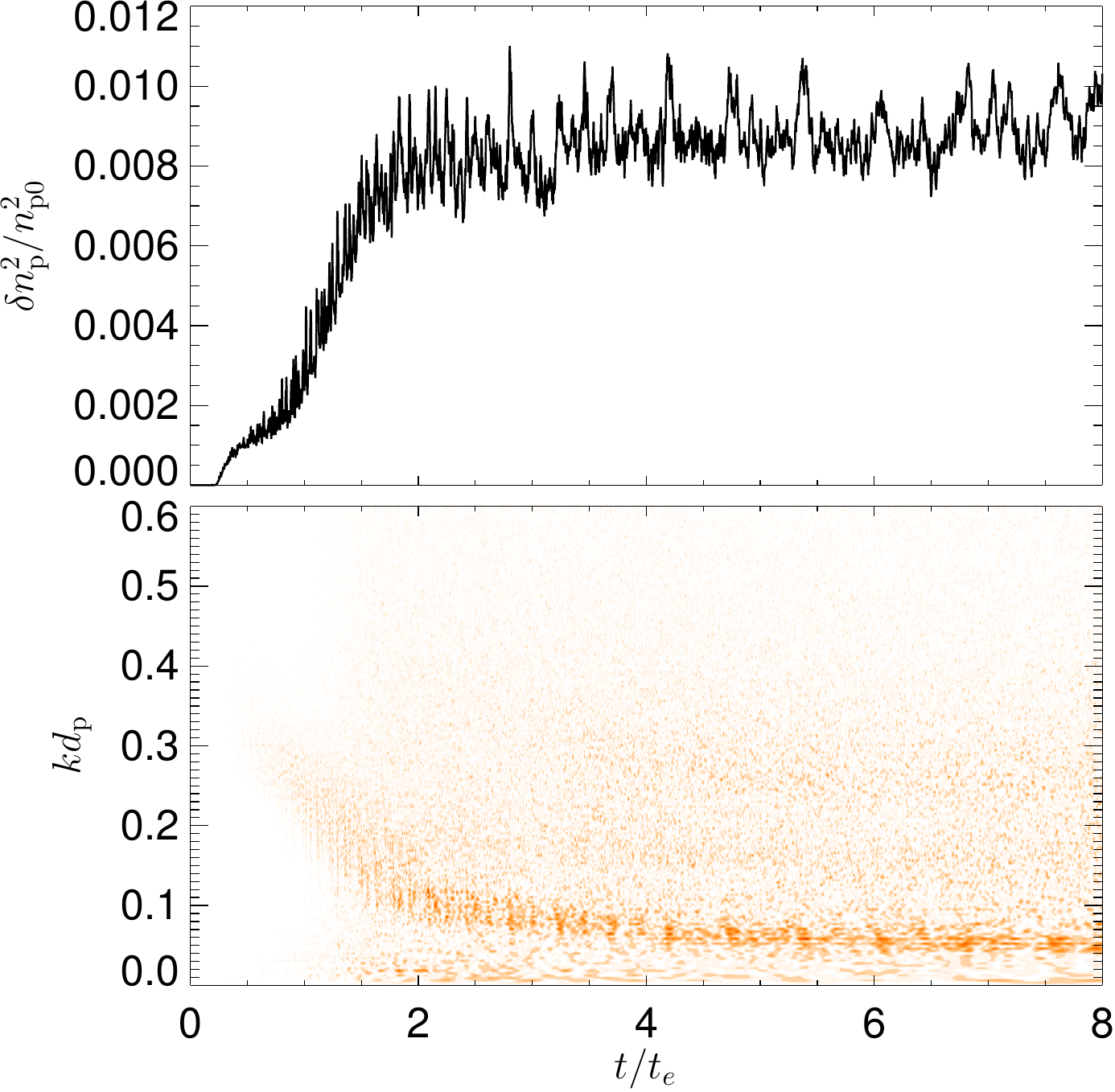}}
\caption{Evolution of the (core) proton density fluctuations:
(top) $\delta n_\mathrm{p}^2$ as a function of $t$; (bottom) color scale plots
of (bottom) $\delta n_\mathrm{p}^2$ as a function of $k$ and $t$.
\label{dn}
}
\end{figure}

The appearance of a secondary population of wave modes 
indicates another, possible secondary instability. The
AIC waves are prone to parametric instabilities
\citep{holl94}; the AIC waves couple to compressible ion acoustic modes 
and AIC waves with different wavelengths. 
Enhanced proton density fluctuations may be an indication of such a process.
Figure~\ref{dn} shows the (core) proton density fluctuations $\delta n_\mathrm{p}^2$
(normalized to the background density $n_{\mathrm{p}0}$)
as a function of time (top panel).
The bottom panel shows the color scale plot
of the fluctuating
magnetic field as a function of time and the wave vector $k$.
Figure~\ref{dn} shows that  weak
density fluctuations appear during the exponential
phase. This is due to the second order effects owing to ponderomotive force.
After the saturation there is an exponential-like growth
of the density fluctuations till $t\sim 1.5t_e$. 
The enhanced density fluctuations appear initially
 around  $k d_\mathrm{p} \sim 0.1$ at $t=2t_e$
and shift to smaller wave vectors at the time goes on.

Combination of Figures~\ref{db} and \ref{dn} suggests
that the AIC waves generated by the pick-up protons
become parametrically unstable and generate the ion acoustic
waves and the secondary population of AIC waves.
An additional, bi-coherence analysis (not shown here) indicate
a weak phase coherence between the two AIC waves and
the compressible modes. The phase coherence starts for $t\gtrsim t_e$
and is kept till $t=8t_e$; it is important to note
 that the situation is symmetric 
with respect to the background magnetic field $\boldsymbol{B}_0$,
the initial AIC waves are generated both parallel and anti-parallel
to $\boldsymbol{B}_0$, the secondary AIC waves as well
as the ion acoustic waves also appear both parallel and anti-parallel
to $\boldsymbol{B}_0$. The presence of forward and backward propagating
modes considerably complicates the phase coherence analysis.
The actual nature of the parametric instability needs more work as
the theoretical properties of parametric instabilities at ion kinetic scales 
are complicated
\cite[cf.,][]{holl94,vasq95}.

\begin{figure}[htb]
\centerline{\includegraphics[width=8cm]{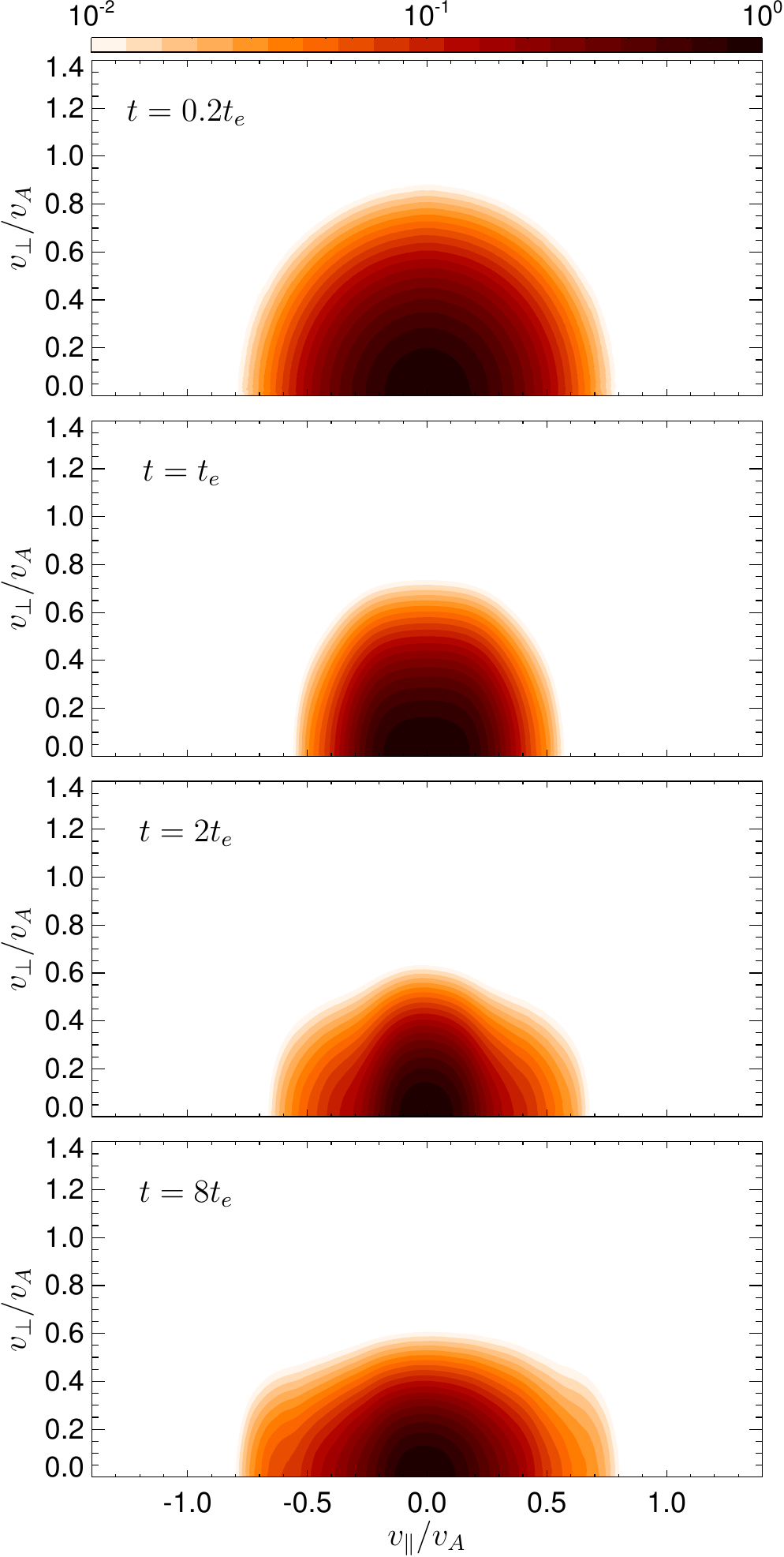}}
\caption{Evolution of the (core) proton velocity distribution function (normalized to
its maximum value) as 
a function of $v_\|$ and $v_\perp$
at $t=0.2t_e$, $t_e$, $2 t_e$, and $8 t_e$ (from top to bottom).
The color scale is shown
on the top.
\label{evolvdfp}
}
\end{figure}

The ion acoustic modes interact through the Landau resonance with
protons and lead to parallel acceleration/heating of the resonant particles \citep{aranal08,mattal10}.
Figure~\ref{evolvdfp} shows the evolution of the core proton velocity distribution functions.
At $t=0.2t_e$, i.e., around the saturation of the initial exponential growth
of the magnetic fluctuations the solar wind protons are essentially unaffected
by the AIC waves.
At $t=t_e$ the solar wind protons exhibit signatures of perpendicular heating (and parallel cooling)
 through the cyclotron 
resonance with the AIC waves (for $|v_\|| \sim 0.3$--$0.6v_A$).
At $t=2t_e$ the solar wind proton distribution functions exhibit regions
with an efficient parallel heating (for $|v_\|| \sim 0.3$--$0.7v_A$) owing to the
Landau resonance with the compressible ion-acoustic waves.

\begin{figure}[htb]
\centerline{\includegraphics[width=8cm]{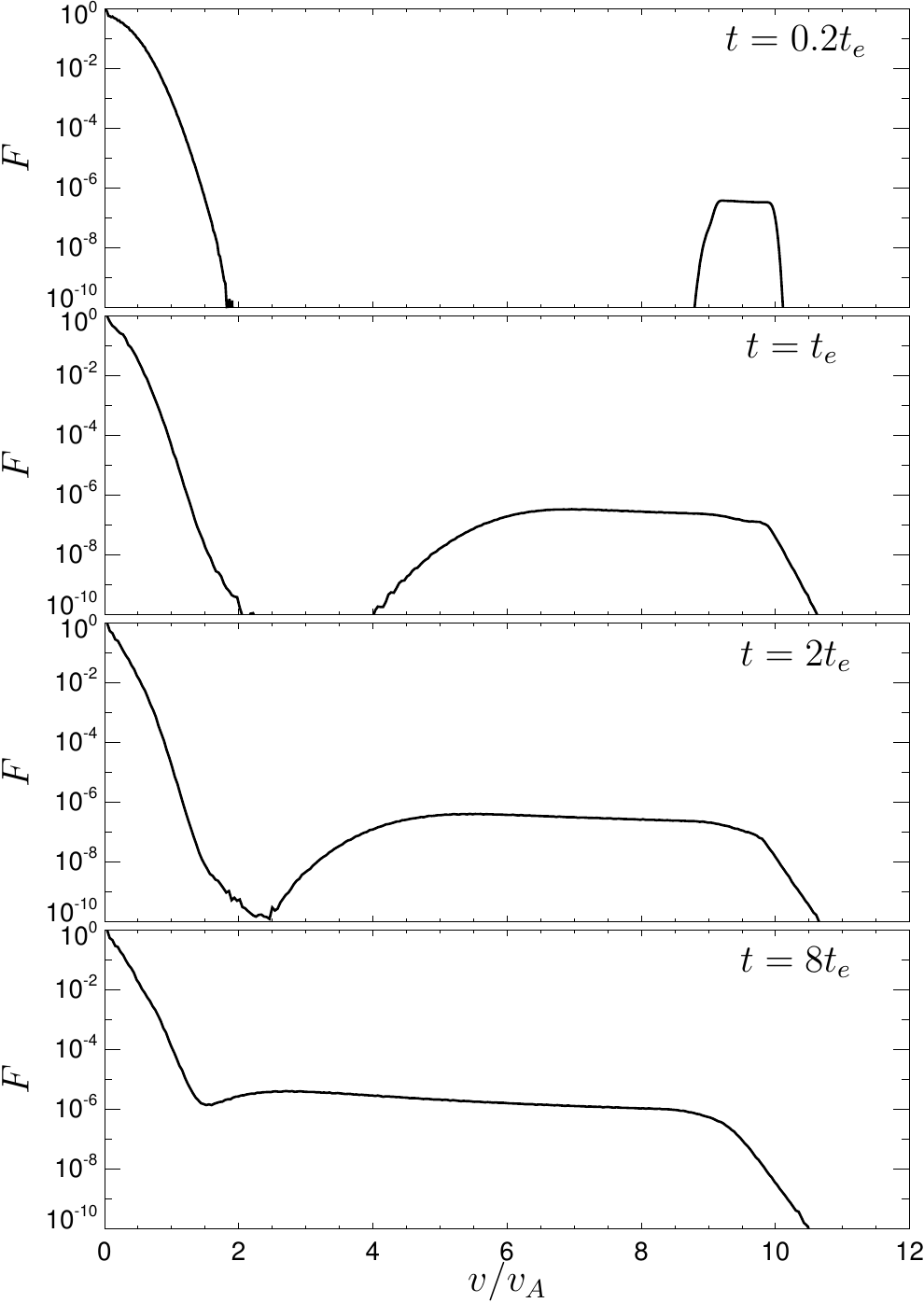}}
\caption{Evolution of the total (core solar wind + pick-up protons)
one-dimensional distribution function as a function of $v$
at $t=0.2t_e$, $t_e$, $2 t_e$, and $8 t_e$ (from top to bottom).
\label{evolfen}
}
\end{figure}

The total proton distribution function consists of both the solar 
wind and pick-up ions. It's interesting to look at this total
distribution function.
Figure~\ref{evolfen} shows the total one-dimensional velocity distribution
function $F$  (normalized to its maximum)
as a function of $v$ at $t=0.2t_e$, $t_e$, $2 t_e$, and $8 t_e$ (from top to bottom). 
The one-dimensional distribution function $F$ 
is obtained by integration of the three-dimensional total proton
distribution function over the spherical angles
 $F=\int (f_\mathrm{p}+ f_\mathrm{pui}) \mathrm{d}\Omega $.
Figure~\ref{evolfen} shows in a more quantitative way 
the thickening of the pick-up ion shell.
At the end of simulation (Figure~\ref{evolfen}, bottom panel)
the one-dimensional distribution function $F$ falls as
$\propto v^{-1.4}$ between 3 and 8 $v_A$. We expect
that the distribution function of pick-ups would eventually become
a power-law like with a cut off at the injection velocity.

\begin{figure}[htb]
\centerline{\includegraphics[width=8cm]{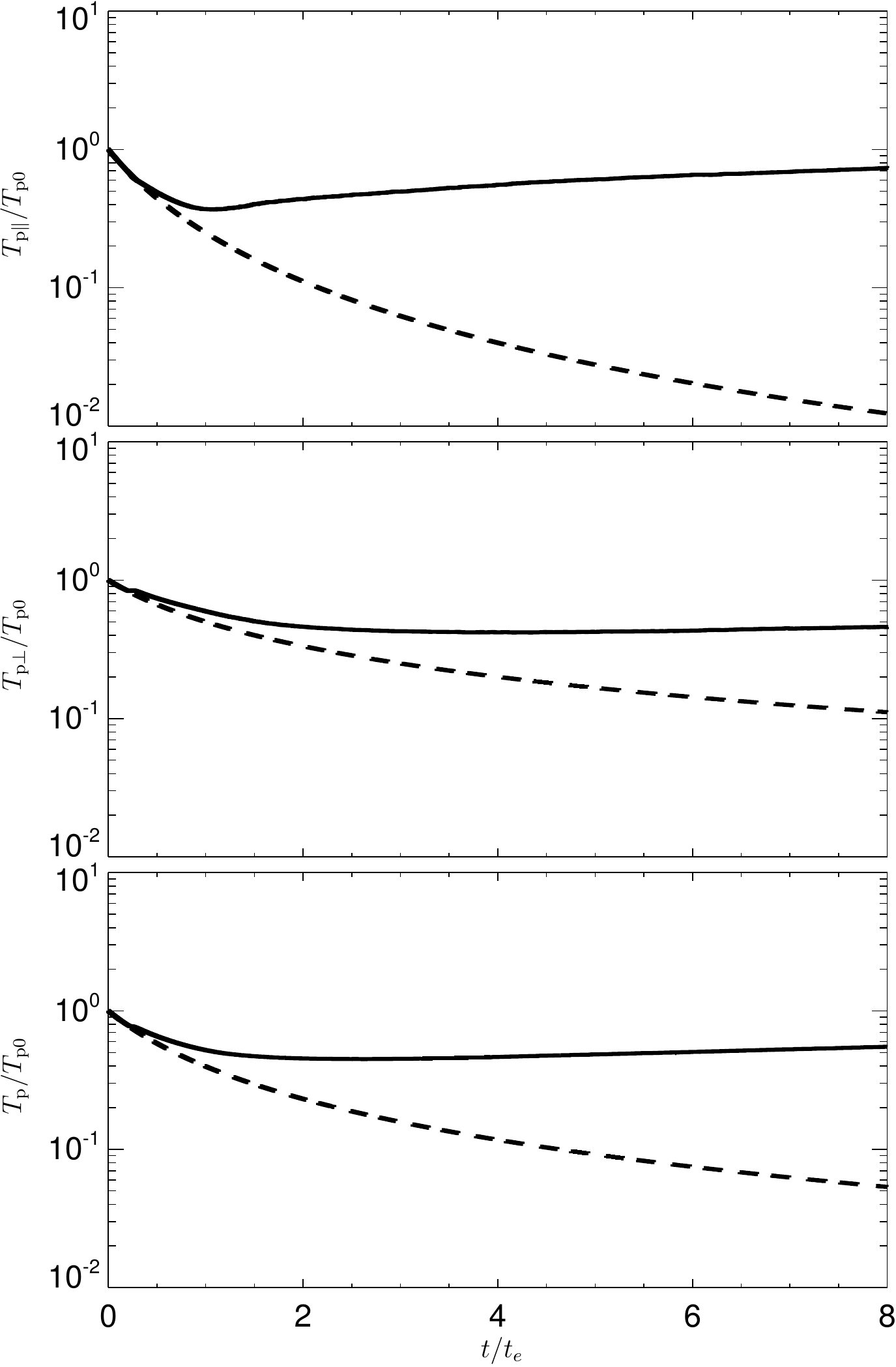}}
\caption{The parallel $T_{\mathrm{p}\|}$ (top), perpendicular $T_{\mathrm{p}\perp}$ (middle),
and total $T_{\mathrm{p}}$ (bottom) temperatures of solar wind protons as functions of time.
The dashed curves show the double-adiabatic prediction
on the the top and middle panels. On the bottom 
panel the dashed curve
 shows the adiabatic prediction.
\label{temp}
}
\end{figure}

Macroscopic evolution of the simulated system is given in 
Figure~\ref{temp} that compares the evolution
of the solar wind proton temperatures (parallel, perpendicular,
and total ones) compared to the double adiabatic/CGL 
prediction \citep{chewal56}
\begin{equation}
\left.T_{\mathrm{p}\|}\right|_{CGL} \propto \frac{n^2}{B^2} \ \ \mathrm{and} \ \
\left.T_{\mathrm{p}\perp}\right|_{CGL}\propto B \label{CGL}
\end{equation}
 (and the adiabatic one $ T_{\mathrm{p}}|_{adiab}\propto n^{2/3}$ for the total temperature).
During the initial exponential growth of the magnetic fluctuations protons 
are cooled in the parallel direction and heated in the perpendicular direction
as expected from the cyclotron resonance between protons and AIC
waves \cite[cf.,][]{hois02}; this cooling is also apparent in Figure~\ref{evolvdfp} (at $t=t_e$).
An efficient parallel heating appears later, $t\gtrsim t_e$, during
the generation of enhanced density fluctuations due to the parametric
instability in agreement of the microscopic properties of the proton velocity distribution function.

\begin{figure}[htb]
\centerline{\includegraphics[width=8cm]{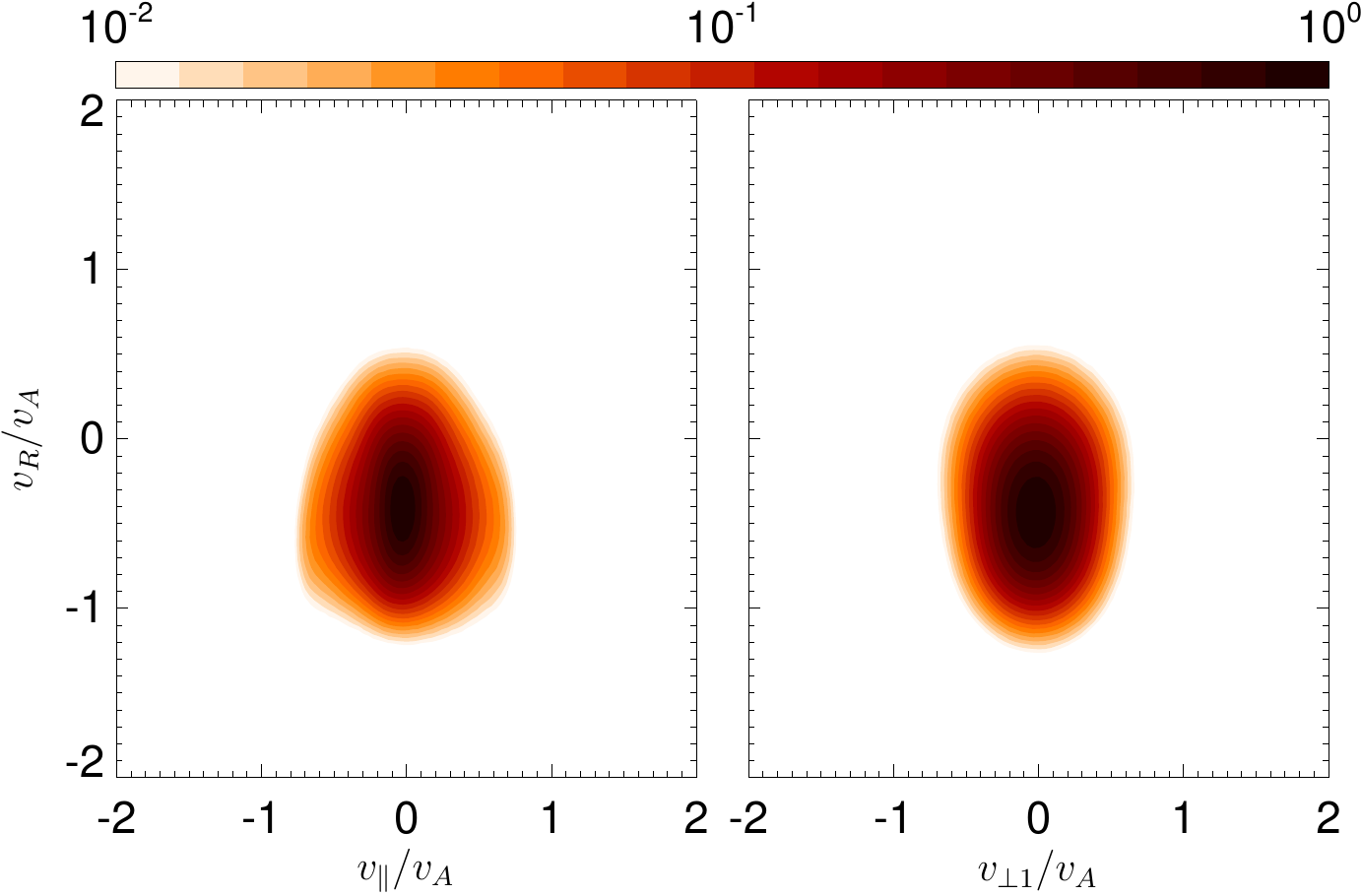}}
\caption{Distribution functions of the hot neutral hydrogen of the solar wind origin at
the end of the simulations: $f_\mathrm{n}$ as a function of $v_\|$ and $v_R$ (left
panel) and  $f_\mathrm{n}$ as a function of $v_{\perp 1}$ and  $v_R$ (normalized to
their maximum value). 
The color scale is shown on the top.
\label{hotn}
}
\end{figure}

Finally, the numerical code keeps information about the velocities
of the neutralized solar wind protons. Figure~\ref{hotn} shows
the velocity distribution function $f_\mathrm{n}$ of the cumulated hot neutrals of the
solar wind origin at the end of the simulation:
The left panel displays $f_\mathrm{n}$ as a function of the radial component
of the velocity $v_R$  and the component parallel to the ambient magnetic field $v_\|$.
The right panels shows $f_\mathrm{n}$ as a function of $v_R$ and the component of
the velocity perpendicular both the radial direction and the ambient magnetic field $v_{\perp 1}$.
The two 2D projections indicate that the 3D velocity distribution function
of hot neutrals is rather complex. The distribution is elongated along
the radial direction owing to the overall deceleration of the solar wind plasma due
to the injection of pick-up protons. The different profiles of the distribution
in $v_\|$ and $v_{\perp 1}$ is a result of generally anisotropic proton velocity
distribution function evolving differently in the direction parallel
and perpendicular to the ambient
magnetic field.

\section{Discussion}
\label{discussion}

We investigated effects of continuous injection of pick-up protons in the distant
solar wind using 1-D hybrid expanding box model. 
We assumed an ideal 1-D homogeneous system parallel to the ambient
magnetic field directed along the transverse direction with
respect to the radial direction. We assumed a slowly expanding ($t_e=10^4\omega_{c\mathrm{p}0}^{-1}$) in the two
transverse directions and a slow continuous injection 
of cold pick-up protons due to the charge-exchange process
with the solar wind protons ($t_{cx}=100 t_e$).
The injection of pick-up protons leads to the formation
of a ring velocity distribution distribution. This distribution
becomes rapidly unstable and generates AIC waves
that scatter through the cyclotron resonance the pick-up protons which 
consequently form a shell velocity distribution distribution.
The AIC waves also scatter the solar wind protons and
heat them in the perpendicular direction and cool them in
the parallel one. 
The continuous injection of pick-up protons keep the instability
active, and, at later times, the AIC waves grow slowly (secularly, about linearly in time).
The AIC waves become eventually unstable with respect
to a parametric instability that leads to formation of a secondary,
shorter wavelength population of AIC waves and
compressible ion-acoustic waves. The ion-acoustic waves interact
with solar wind protons through the Landau resonance. The combined effect
of the AIC and ion-acoustic waves lead to
an efficient proton parallel and perpendicular heating.
The pick-up proton shell distribution thickens during the evolution
due to diffusion owing to the AIC waves and
due to the expansion driven cooling; eventually, a power-law
distribution is expected with a cut off near the injection velocity. 
The pick-up proton generated AIC waves and the ion-acoustic waves generated by the parametric instabilities  may be partly
responsible for the enhanced level of density fluctuations observed in the outer heliosphere
\citep{bellal05,zankal12}.

In the presented simulation we initialized the solar wind protons
with $\beta_\mathrm{p}=0.2$. To test the sensitivity of the
simulation results with respect to $\beta_\mathrm{p}$
we performed additional simulations with
$\beta_\mathrm{p}=0.1$ and 1.
For higher solar wind proton beta the cyclotron perpendicular heating becomes
less efficient but overall behavior remain the same. For higher $\beta_\mathrm{p}$
one expects that the mirror instability would become important \citep{garyal97} and, also, properties
of the parametric instabilities depend on  $\beta_\mathrm{p}$ \citep{holl94}. Therefore, the present
results are relevant for $\beta_\mathrm{p}\lesssim 1$.

The present model is in many respects simplified, only one dimension is considered, the solar wind velocity is assumed
to be constant and is about ten time faster than in the reality. Also, the system is assumed 
homogeneous, no pre-existing fluctuations/turbulence is assumed, Coulomb collisions
and electron impact ionization are neglected, etc.
However,
the model self-consistently describes the kinetic plasma behavior
in the expanding solar wind where pick-up protons are continuously injected.

The one-dimensional geometry strongly reduces the available physics,
only parallel propagating modes are allowed. In a low beta
plasma the ring is expected to generate cyclotron waves  primarily
along the magnetic field which justifies the 1-D geometry but
other instabilities (such as
the mirror instability) may appear at oblique angles 
and may modify the nonlinear behavior.
Also the 1-D geometry tend to leads to larger amplitude fluctuations
the are prone to parametric instabilities. At 2-D and 3-D the
effect of parametric instabilities will be likely reduced.

The numerical resolution of the code could be source of other problems;
our choice of the spatial grid, the box size, and the
time step guaranties a good resolution of the AIC waves. While the used
number of particles per cell is substantial (note that at $t=t_e$ there are about
$10^3$ particles per cell for pickup protons) the resulting numerical noise 
may lead to enhanced scattering of the ring pick-up protons.
 \citep{floral16}. This is likely a minor
problem since the continuous injection of pickup ions tends to keep the system
unstable and rapidly generates fluctuations well above the numerical noise level.

The presence of large scale fluctuations/turbulence will modify the
initial local pick-up ion distribution function and
the resulting instabilities \citep{zaca00}. For injection at nearly parallel
angles with respect to the magnetic field beam-type instabilities are expected 
\citep{gary93}. These instabilities have different nonlinear properties but quite
generally one expects formation of partial spherical shells
  \cite[cf.,][]{wiza94,mattal15}.

Coulomb collisions are expected to be weak in the solar wind but may lead at the expansion time scale
to scattering of pick-up ions \cite[via interaction with the solar wind ions, cf.,][]{tracal15,hell16}
 reducing thus the source of free energy for instabilities.
Electron impact ionization is expected to have generally a subdominant effect with respect
to the charge-exchange process in the solar wind \cite[but it is likely important in the heliosheath, cf.,][]{deckal08}.
This process would have a similar effect as the charge exchange (i.e., generation of
pick-up ions) but without the reduction of the solar wind ion density. We expect the
overall effect of electron impact ionization would comparable to that of charge exchange.

The connection
between instabilities driven by the pick-up ions and turbulence
remains an important open question. The turbulent fluctuations (or
another wave activity)
already present in the solar wind would scatter the pick-up ions
and possibly reduce the source of free energy for instabilities. 
On the other hand, instabilities driven by pick-up ion ring-beam 
velocity distribution function generate typically waves at
short, ion scales, often at nearly parallel angles with
respect to the ambient magnetic field. Coupling between
the strongly oblique turbulent fluctuations and quasi-parallel
waves at ion scales is likely weak
\cite[while waves generated at strongly oblique angles
seem to participate in the cascade, cf.,][]{hellal15}.
 There are many open problems that will be subject of
future work. 

In concluding,
the present work shows that
 (i) collisionless plasma with energetically
important population of pick-up ions generally needs a fully kinetic treatment,
 (ii) pick-up ion generated waves are able to directly quite efficiently heat the 
solar wind protons, 
(iii) the distribution function of pick-up ions at
later times/larger distances has a wide spread of velocities/energies
owing to scattering of the initial distribution on the generated waves
and to the expansion-driven cooling,
(iv) the hot neutrals of the solar wind origin have a complex velocity
distribution function since they are neutralized at different distances 
and the solar wind ion temperature and the bulk velocity vary substantially
over the time/distance.

\end{document}